\documentclass[a4paper,12pt]{article}

\date{}

\usepackage[dvips]{graphicx}
\usepackage[cp 1251]{inputenc}
\usepackage{amsmath}
\usepackage{citesort}

\numberwithin{equation}{section}

\reversemarginpar

\title{Photoproduction of the $\pi N$ pair on nuclei and isobar configurations}
\author{Glavanakov I. V., Tabachenko A. N.\\
Institute of Physics and Technology, Tomsk Polytechnical  University, \\ Tomsk, Russia}

\begin{document}

\maketitle

\begin{center}
\noindent
\parbox{15cm}
{\small
A model of $\pi N$ pair photoproduction on nuclei at high momentum transfer is presented.
The $A(\gamma,\pi N)B$ reaction amplitude is obtained by means of an extended impulse approximation,
according to which the nuclear wavefunction includes delta-isobar components in addition to nucleon
components. A one-particle transition operator is defined in terms of the two-body $\gamma N\to\pi N$
and $\gamma \Delta\to\pi N$ photoproduction amplitudes. Direct and exchange mechanisms of the
nuclear photoproduction reaction are studied, and numerical estimates are made and presented of
$^{12}{\rm C}(\gamma,\pi^- p){^{11}{\rm C}}$ and $^{12}{\rm C}(\gamma,\pi^+ p){^{11}{\rm Be}}$
differential cross sections at photon incident energies in the $\Delta(1232)$ resonance region.
}\\
\end{center}

\section{Introduction}

The photoproduction of pion on the nucleus, which is accompanied by
emission of nucleons, is useful for the study of such problem of the nuclear physics
as "$\Delta$-isobar in nuclei". Usually three substantially different facets of
the delta-isobar are considered in the nucleus. These facets differ in the isobar-production
mechanism and isobar state. Quasifree isobar production in a "free" state
in the scattering of high-energy particles on nuclei has received the most comprehensive
study. In this case, the isobar is produced nearly on-mass-shell, because the energy-momentum
transfer to a bound intranuclear nucleon is quite high. Such an isobar
propagates in the nucleus involved, interacting with the closest nucleons,
and decays with a high probability to a pion and a nucleon or undergoes the transition
$\Delta N\rightarrow NN$ accompanied by the knockout of two nucleons.

Another facet of $\Delta$-nucleus physics is associated with isobar
configurations in the ground state of the nuclei. At intermediate distances
the most of the attraction between two nucleons of nuclei comes from the exchange of
 two pions, between which one or two $\Delta$s can be created. Thus, as a result of
nucleon collisions  the creation of the virtual delta-isobars is possible.
These delta-isobars are produced far from the mass shell and therefore cannot
undergo the decay $\Delta \rightarrow N \pi$, but they can transit into
a free state upon the transfer of the required 4-momentum to them from a high-energy particle.
The excitations of the bound nucleons are most intensive at the large momentum transfer,
therefore the virtual delta-isobars are connected with the high-momentum components
of nuclear wave function. In the framework of the non-relativistic
semi-phenomenological model the virtual isobars have led to the so-called
isobar configurations in nuclei \cite {1,2}. In this model the conventional
wave function consisting of nucleons is supplemented  by exotic
components in which one or several nucleons are internally excited, i.e.
are baryon resonances or isobars.

The third facet, which has to be studied adequately, is associated with a
hypothetic quasibound delta-nucleus state of a nucleus. In many respects
delta-nucleon interaction is similar to nucleon-nucleon interaction,
which is attractive  \cite {3}. Therefore, it can be hypothesized that under favorable
conditions such that the momentum of the product or knock-on isobar is small in
relation to the momentum of nucleons  bound in the nucleus involved,
the delta isobar and the residual nucleus may form a highly excited bound state ($\Delta$-nucleus).
This is not an ordinary bound nuclear state, since it is unstable with respect
to the emission of a pion or a pion-nucleon pair. Moreover, the lifetime of a free
isobar is substantially shorter than the time required for the formation of a normal
collective nuclear state, according to \cite {4}, therefore, we will refer to the states in question
as quasibound states.

The possible existence of bound and resonance  delta-nucleon and delta-nucleus states
was widely discussed in \cite {5,6,7,8,9}. From the experimental point of view, conclusion of the
work \cite {4} that a quasibound delta-nucleon state of isospin $T=2$ and spin-parity $J^{P}=2^{+}$
may exist is the most appealing. The binding energy of this state is estimated
at 10 to 40 MeV depending on the approximations used in relevant calculations.
Last time this problem was considered in the works \cite {10,11}, where the results of the
experiment at  Tomsk synchrotron were discussed. The cross
section for the $^{12}$C$(\gamma,\pi^{-}p)$ reaction was measured in the $\Delta(1232)$-resonance region.
This experiment possibly indicates  the existence of quasibound isobar-nucleus states. The analogical
conclusions were made in the works \cite {12,13}, in which  the author considered data from three
experiments performed at the linear accelerator in Saclay \cite {14}, at Tomsk synchrotron \cite {15} and
at  MAMI accelerator in Mainz \cite {16} and devoted to exploring the photoproduction of single pions on
light nuclei that is accompanied by nucleon emission.

The conclusions of the works \cite {10,11,12,13} were based on comparison of the experimental
data of the $A(\gamma,\pi N)$ reaction
with the theoretical predictions obtained in the frame of the model,
based on the hypotheses of the $\Delta$-nucleus existence.
This reaction mechanism is manifested in the region of high momentum transfers to the
residual nuclear system. However, in the same kinematical region other possible  concomitant
mechanisms of the reaction also occur. Particularly, manifestations of the
isobar configurations in the nucleus ground state and meson-exchange currents are possible.
An analysis of these reaction mechanisms is needed for testing conclusions drawn in \cite {10,11,12,13}
about the existence of quasibound isobar-nucleus states and for interpretation of
$(\gamma,\pi^+ p)$ and $(\gamma,\pi^- n)$ reaction data.

In this work we study the influence of isobar configurations on
photoproduction of pion-nucleon pairs on light nuclei with  closed shells.
The basic ingredients of the reaction model presented are nucleus density matrices, taking into account
nucleon and isobar degrees of freedom, and single-particle operators of  $\gamma N\rightarrow\pi N$
and $\gamma \Delta\rightarrow\pi N$ transitions. Direct and exchange reaction mechanisms are considered.
Using this model, we calculate the contribution of isobar configurations
to the  cross section of the  $^{12}$C$( {\gamma ,^{}\pi ^{-} p})^{11}$C reaction in the area where the
existence of quasibound isobar-nucleus states is expected and estimate numerically the cross section of the
$^{12}$C$({\gamma ,\pi ^{ +} p})^{11}$Be reaction in the region of high momentum transfer.

\section{Amplitude of the $ A(\gamma ,\pi N)B$ reaction}

The matrix element  of the $S$-matrix between the initial state {\it i} and the final state {\it f}, describing
the reaction of the pion photoproduction on nucleus {\it A} is accompanied by the emission of nucleon {\it N}
and the formation of a residual nucleus {\it B} can be represented in the form
\begin{equation}\label{eq:2.1}
<f \mid S \mid i>=- 2\pi\, \delta(E_{\gamma}+ E_{T}-E_{\pi}-E_{N}-E_{R})
\frac{T_{fi}}
{(2E_{\gamma} 2E_{\pi})^{1/2}},
\end{equation}
where $E_{\gamma}$, $E_{T}$, $E_{\pi}$, $E_{N}$ and $E_{R}$ are energies of the photon, the initial nucleus {\it A},
the pion, the nucleon and the residual nucleus {\it B};  $T_{fi}$ is the transition matrix element of the
$A(\gamma ,\pi N)B$ reaction.

For description of nuclei we will use the approach
developed in the works of Arenhovel et. al. \cite{2,17,18} for
study of the isobar configurations in the ground states of the light nuclei. Here, we
apply this formalism for the description of the  nuclear reactions.

 According to \cite{2,17,18}, baryons bound in the nucleus, in addition to the space $\mathbf{r}$, spin {\it s}, and
isospin {\it t} coordinates ( $\mathbf{r},s,t\equiv x$), are characterized also by the intrinsic
coordinate {\it m} $(x,m\equiv X)$. An eigenfunction $\Psi_{\beta}(X_{1},...,
X_{A})$ of  hamiltonian {\it H} of the system of {\it A} particles  with eigenvalue $E_{\beta}$
is a superposition of the wave functions concerned with  different intrinsic configurations
$$
\Psi_{\beta}(X_{1},...,X_{A})=\sum_{n} \Psi_{\beta}^{n} (X_{1},...,X_{A}),
$$
$$
\Psi_{\beta}^{n} (X_{1},...,X_{A})=A_{n}\phi_{n}(m_{1},...,m_{A}) \, \psi _{\beta}^{n} (x_{1},...,x_{A}).
$$
Here $\phi_{n} (m_{1},...,m_{A})$ is the intrinsic wave
function of {\it A} particles.  The index $n\equiv n_{1},...,n_{A}$ characterizes
the intrinsic state of the particles. For instance, the state index describing the intrinsic configuration of the
nucleon system is written as $n\,=\,N_1,N_2,...,N_A$; if the first particle is in the state of
isobar, but the rest are nucleons, the intrinsic state index is written as $n\,=\,\Delta _1,N_2,...,N_A$.
By definition, $\psi _{\beta}^{n} (x_{1},...,x_{A})$ is the wave function describing the state
of {\it A} particles  with quantum numbers
$\beta\equiv \beta_{1},...,\beta_{A}$ in the usual space, spin and
isospin spaces, and with quantum numbers {\it $n\,=\,n_{1},...,n_{A}$} in the intrinsic space.
The wave function $\psi _{\beta}^{n} (x_{1},...,x_{A})$
should be antisymmetric for particles in the same intrinsic state. The remaining
antisymmetrization for particles in different intrinsic states is done by the operator
$A_{n}$.

 In the frame of this approach we define the matrix element $T_{fi}$
 of the $A(\gamma ,\pi N)B$ reaction in configuration space, which, in addition
 to the usual space, spin, isospin coordinates, also includes the  intrinsic coordinates, as
$$
{T_{fi}}=
A\int d(X'_{1},X_{1},...,X_{A})\Psi^{\ast}_{F}(X'_{1},X_{2},...,X_{A}) \times
$$
$$
\times <X'_{1}\mid t_{\gamma \pi}\mid X_{1}>\Psi_{T}(X_{1},...,X_{A}).
$$
Here the integral sign denotes the integration over the space variables and
 summation over the spin, isospin and intrinsic variables;
$\Psi_{F}(X_{1},...,X_{A})$ is the antisymmetrized wave function of the final nuclear
system {\it F} including residual nucleus {\it B} and nucleon {\it N} in the free state;
 $\Psi_{T}(X_{1},...,X_{A})$ is the antisymmetrized wave function of
 nucleus {\it A}; $t_{\gamma \pi}$ is the single-particle operator of the pion photoproduction on
a free baryon.

Let us present the wave function $\Psi_{F}$ of the final nuclear system as the antisymmetrized product of the
wave function $ \varphi_{\textbf{p}_n}$ of the free nucleon with the momentum $\textbf{p}_n$
and the wave function $\Psi_{f}$ of nucleus {\it B} in the state {\it f}:
$$
\Psi_{F}(X_{1},...,X_{A})=A^{p}_{1;2...A}\varphi_{\textbf{p}_n}(X_{1})\Psi_{f}(X_{2},...,X_{A}),
$$
where
$ A^{p}_{1;2...A}$
is the antisymmetrization operator.
Then,  we obtain for the {\it T}-matrix
$$
T_{fi}=T_{d}-T_{e}.
$$
Here the direct amplitude  $T_{d}$ is
$$
T_{d}=\sqrt{A}\int d(X'_{1},X_{1},...,X_{A})\varphi^{*}_{\textbf {p}_n}(X'_{1})
\Psi^{*}_{f}(X_{2},...,X_{A})\times
$$
$$
\times <X'_{1}|t_{\gamma \pi}|X_{1}>\Psi_{T}(X_{1},...,X_{A}),
$$
in which the ''active'' particle with number 1 interacts with the photon and
changes the state from bound to free;
the exchange amplitude $T_{e}$ is
$$
T_{e}=\sqrt{A}(A-1)\int d(X'_{1},X_{1},...,X_{A})\varphi^{*}_{\textbf{p}_n}(X_{2})
\Psi^{*}_{f}(X'_{1},X_{3},...,X_{A})\times
$$
\begin{equation}\label{eq:2.5}
\times <X'_{1}|t_{\gamma \pi}|X_{1}>\Psi_{T}(X_{1},...,X_{A}),
\end{equation}
in which the ''active''  particle remains in
the bound state after the interaction with the photon.

Let us now write the square of the modulus of the amplitude $ T_{fi}$
$$
T_{fi}T^{*}_{fi}=T_{d}T^{*}_{d}+T_{e}T^{*}_{e}-T_{d}T^{*}_{e}-T_{e}T^{*}_{d}
$$
The  square of the modulus of the direct amplitude $ T_{d}$ is
$$
T_{d}T^{*}_{d}=A\int d(X^{'}_{1},X_{1},\widetilde{X}{^{'}_{1}}\widetilde{X}{_{1}})
\varphi^{*}_{\textbf {p}_n}(X'_{1})<X'_{1}|t_{\gamma \pi}|X_{1}>\psi _{A-1,A}(X_{1})\times
$$
$$
\times \psi^{*}_{A-1,A}(\widetilde{X}{_{1}})<\widetilde{X}{_{1}}|t^{+}_{\gamma \pi}|\widetilde{X}{'}_{1}>
\varphi_{\textbf {p}_n}(\widetilde{X}{'_{1}}),
$$
where
$$
\psi _{A-1,A}(X_{1})=\int d(X_{2},...,X_{A})\Psi^{*}_{f}(X_{2},...,X_{A})\Psi_{T}(X_{1},...,X_{A})
$$
is the overlap function.

 The differential cross section of the reaction
$A(\gamma,\pi N)B$, summed over all  the final states of the
nucleon and the residual nucleus  will be considered. Let us accept the condition of the completeness of final states
$$
\sum_{f}\Psi^{*}_{f}(X_{2},...,X_{A})\Psi_{f}({\widetilde{X}}_{2},...,{\widetilde{X}}_{A})=
\delta (X_{2}-\widetilde{X}_{2})\,...\,\delta (X_{A}-\widetilde{X}_{A}),
$$
where sum is taken over all  the final states of the residual nucleus.
In this case the expression for the square of the modulus of the amplitude $T_{d}$ is
$$
\sum _{f}T_{d}T^{*}_{d}=A\int d(X'_{1},X_{1},\widetilde{X}'_{1},\widetilde{X}_{1})\,
\varphi^{*}_{\textbf {p}_n}(X'_{1})<X'_{1}|t_{\gamma \pi}|X_{1}> \times
$$
$$\times \,\rho(X_{1};\widetilde{X}_{1})<\widetilde{X}{_{1}}|t^{+}_{\gamma \pi}|\widetilde{X}{'}_{1}>
\varphi_{\textbf{p}_n}(\widetilde{X}{'_{1}}),
$$
where
$$
\rho(X_{1};\widetilde{X}_{1})=\int d(X_{2},...,X_{A})\Psi_{T}(X_{1},...,X_{A})
\Psi^{*}_{T}(\widetilde{X}_{1},X_{2},...,X_{A})
$$
is the one-body density matrix.

We will now consider the exchange amplitude $T_{e}$.
In the case of the  exchange mechanism of the charge pion photoproduction,
the ''active'' nucleon remaining in the bound state
can transit to the states, which are above or low the Fermi level.
The last vacant levels were produced as the result of the process of the $\Delta$-isobar production
by means of the transition $NN\rightarrow \Delta N $. Also, the ''active'' nucleon
can transit to the vacant level arose as the result of
the virtual decay $A\rightarrow (A-1)+N$. In the case  of the neutral pion photoproduction
the  exchange amplitude contains additionally the transitions
$\gamma N\rightarrow N\pi $  without change of the nucleon state.

In the case, if the ''active'' nucleon goes to the state, which is above the Fermi level,
the wave function of the residual nucleus
may be written as
\begin{equation}\label{eq:2.8}
\Psi_{f}(X'_{1},X_{3},...,X_{A})=A^{S}_{u;1...A\neq k...l}\, \Psi_{\beta_{u}}(X'_{1})\,
\Psi_{(\beta_{k}...\beta_{l})^{-1}}(X_{3},...,X_{A}),
\end{equation}
where $\Psi_ {\beta}(X)$ is one-particle wave function of the  nucleon bound in nuclei,
$\beta _{u}$ is the index of the nucleon state which is above the Fermi level,
$(\beta_{k}...\beta_{l})^{-1}$ is the hole state of the bound system of baryons with
numbers 3, ..., {\it A},
the antisymmetrization operator $A^{S}_{u;1...A\neq k...l}$ rearranges the indices of the
nucleon states.

As all nucleons of the wave function $\Psi_{T }$ are lower than the  Fermi level,
the nonzero contribution of the exchange amplitude arises from the first summand
$$
\Psi_{\beta_{u}}(X'_{1})\, \Psi_{(\beta_{k}...\beta_{l})^{-1}}(X_{3},...,X_{A})
$$
of the expression ($\ref{eq:2.8}$). As a result, the  square of the modulus of the exchange amplitude is
$$
\displaylines{
T_{e}T^{*}_{e}=A(A-1)\int d(X'_{1},X_{1},X_{2},\widetilde{X}'_{1},\widetilde{X}_{1},\widetilde{X}{_{2}})
\varphi^{*}_{\textbf{p}_n}(X_{2})\,\Psi^{*}_{\beta_{u}}(X'_{1})<X'_{1}|t_{\gamma \pi}|X_{1}>
\cr
\times\,
\psi _{A-2,A}(X_{1},X_{2})\,\psi^{*}_{A-2,A}(\widetilde{X}_{1},\widetilde{X}_{2})
<\widetilde{X}{_{1}}|t^{+}_{\gamma \pi}|\widetilde{X}'_{1}>
\Psi_{\beta_{u}}({\widetilde{X}}'_{1})
\varphi_{\textbf{p}_n}(\widetilde{X}_{2}),
\cr}
$$
where
$$
\psi _{A-2,A}(X_{1},X_{2})=\int d(X_{3},...,X_{A})\Psi^{*}_{(\beta_{k}...\beta_{l})^{-1}}(X_{3},...,X_{A})
\Psi _{T}(X_{1},...,X_{A}).
$$

If the set of the states $(\beta_{k}...\beta_{l})^{-1}$ is complete, then
$$
\sum _{f}T_{e}T^{*}_{e}= A(A-1)\,\sum _{u}\int d(X'_{1},X_{1},X_{2},\widetilde{X}'_{1},\widetilde{X}_{1},\widetilde{X}{_{2}})
\varphi^{*}_{\textbf{p}_n}(X_{2})\Psi^{*}_{\beta_{u}}(X'_{1})\,\times
$$
$$
\times<X'_{1}|t_{\gamma \pi}|X_{1}>
\rho(X_{1},X_{2};\widetilde{X}_{1},\widetilde{X}_{2})
<\widetilde{X}{_{1}}|t^{+}_{\gamma \pi}|\widetilde{X}'_{1}>
\Psi_{\beta_{u}}({\widetilde{X}}'_{1})
\varphi_{\textbf{p}_n}(\widetilde{X}_{2}).
$$
Here
$$
\rho(X_{1},X_{2};\widetilde{X}_{1},\widetilde{X}_{2})=\int d(X_{3},...,X_{A})\,\Psi_{T}(X_{1},X_{2},...,X_{A})\,
\Psi^{*}_{T}(\widetilde{X}_{1},\widetilde{X}_{2},...,X_{A})
$$
is the two-body density matrix.

We will neglect the contribution of the products $T_{d}T^{*}_{e}$ and $T_{e}T^{*}_{d}$, as
the kinematical regions, in which the basic contribution of the direct and exchange amplitudes
 in cross section differ considerably.

 \section{Nucleus wave function }

The nucleus wave functions $\Psi _{\beta}(X_{1},...,X_{A})$
 satisfy the following Schr$\ddot{o}$dinger equation
\begin{equation}\label{eq:3.1}
 ( H- E_{\beta})\,\Psi _{\beta}(X_{1},...,X_{A})=0,
\end{equation}
 where the hamiltonian of the system {\it H} acts on spatial, spin, isospin, and intrinsic coordinates.
According to \cite{18} the hamiltonian {\it H}  has the form
$$
  H=\sum_{i=1}^{A}\left(T_i+I_i\right)+\sum_{i<j}V_{ij}\,=\,T\,+I\,+\,V.
$$
 Here $T_i$ is the kinetic energy operator of {\it i}-particle, $I_i$ is the part
connected with the intrinsic degrees of freedom,  $V_{ij}$ is the
two-particle interaction. The operators {\it T} and {\it V}, unlike those of standard nuclear
physics, also depend on the intrinsic degrees of freedom. The operators {\it T} and
{\it I} are diagonal by the intrinsic degrees of freedom.

The wave function of the  nucleus in eq. (\ref{eq:3.1}) may be written as
\begin{equation}\label{eq:3.2}
\Psi(X_{1},...,X_{A}) = \Psi _{N}(X_{1},...,X_{A})+\Psi _{\Delta}(X_{1},...,X_{A}),
\end{equation}
where
$$
\Psi _{N}(X_{1},...,X_{A})=\phi _{N}(m_{1},...,m_{A})\psi^{N}_{\beta}(x_{1},...,x_{A})
$$
is the wave function of the  nucleus in the state, when all particles of the nucleus
are nucleons; intrinsic state index $N \equiv N_1,N_2,...,N_A$;
$$
\Psi _{\Delta}(X_{1},...,X_{A}) = \sum\limits _{\mathnormal{\Delta}} A_{\mathnormal{\Delta}}\,
\phi _{\mathnormal{\Delta}}(m_{1},...,m_{A})\psi^{\mathnormal{\Delta}}_{\beta}(x_{1},...,x_{A})
$$
is the wave function of the  nucleus, which includes the states with one isobar
$\mathnormal{\Delta}\,=\,\Delta _1, N_2,...,N_A$, two isobars
$\mathnormal{\Delta}\,=\,\Delta _1, \Delta _2, N_3,...,N_A$ and etc.  The wave functions
$\Psi _{N}$ and $\Psi _{\Delta}$ are normalized correspondingly by $N_N$ and $N_\Delta$.

The wave functions $\Psi _{\beta}^{\mathnormal{\Delta}}(X_{1},...,X_{A})$ of the
$\mathnormal{\Delta}$-configuration satisfy
the following equation
\begin{equation}\label{eq:3.3}
( H- E_{\beta})\,\Psi_{\beta}^{\mathnormal{\Delta}}(X_{1},...,X_{A})=
-\sum _{n'\neq \mathnormal{\Delta}} V\,\Psi _{\beta}^{n'}(X_{1},...,X_{A}).
\end{equation}

Since those configurations, in which one or several nucleons are in an intrinsically excited state,
are expected to be small because of the large excitation energy, for
this equation one can  find an approximation solution  in a perturbative approach,
according to which
one can leave only $ n'= N= N_1,N_2,...,N_A$ configuration on the right-hand side of eq. (\ref{eq:3.3}).
Then, in this approximation we will have the equation
$$
( H- E_{\beta})\,\Psi_{\beta}^{\mathnormal{\Delta}}(X_{1},...,X_{A})=
-  V\,\Psi _{\beta}^{N}(X_{1},...,X_{A}).
$$

In our model we will take into account only the dominant one-$\Delta$ configuration.
Assuming that only two nucleons are involved in the excitation of the nucleon internal degrees
of freedom, wave function $\Psi_{\Delta}$ of one-$\Delta$ configuration can be written as the
superposition of the products of the wave function $\Psi^{\Delta N}_{[\beta_{i}\beta_{j}]}$
of $\Delta N$ system, which includes an isobar and the second nucleon (the participant of the transition
$NN \rightarrow\Delta N$) and the wave function $\Psi^{N}_{(\beta_{i}\beta_{j})^{-1}}$ describing
the state of the nucleon core, which includes other A--2 nucleons,
\begin{equation}\label{eq:3.5}
\Psi _{\Delta}(X_{1},...,X_{A})=A^{P}_{12;3,...,A}\sum_{ij}\Psi^{\Delta N}_{[\beta_{i}\beta_{j}]}(X_{1},X_{2})
\Psi^{N}_{(\beta_{i}\beta_{j})^{-1}}(X_{3},...,X_{A}).
\end{equation}
Here
$$
A^{P}_{12;3,...,A}=\sqrt{\frac{2}{A(A-1)}}\left(1-\sum^{A}_{i=3}(P^{P}_{1i}+P^{P}_{2i})+
\sum^{A-1}_{i=3}\sum^{A}_{j=i+1}P^{P}_{1i}P^{P}_{2j}\right)
$$
is the antisymmetrizaton operator, the operator $P_{ik}^P$ interchanges the {\it i}-th and {\it k}-th nucleons,
\begin{equation}\label{eq:3.5.1}
\Psi^{\Delta N}_{[\beta_{i}\beta_{j}]}(X_{1},X_{2})=A^{P}_{1;2}\phi_{\Delta N}(m_{1},m_{2})
\psi^{\Delta N}_{[\beta_{i}\beta_{j}]}(x_{1},x_{2}),\ A^{P}_{1;2}=\frac{1}{\sqrt{2}}(1-P^{P}_{12}),
\end{equation}
$$\Psi^{N}_{(\beta_{i}\beta_{j})^{-1}}(X_{3},...,X_{A})=\phi_{N}(m_{3},...,m_{A})
\psi^{N}_{(\beta_{i}\beta_{j})^{-1}}(x_{3},...,x_{A}).
$$

The wave function $\psi_{\beta}^{\mathnormal{\Delta}}$  of {\it A}
 particles with quantum number $\mathnormal{\Delta}$ satisfy the following Schr$\ddot{o}$dinger equation
\begin{equation}\label{eq:3.6}
 (\phi_{\mathnormal{\Delta}},( H- E_{\beta})A_{\mathnormal{\Delta}}\phi_{\mathnormal{\Delta}})
 \psi_{\beta}^{\mathnormal{\Delta}}(x_{1},...,x_{A})=
-(\phi_{\mathnormal{\Delta}}, V\phi_{N})\psi_{\beta}^{N}(x_{1},...,x_{A}),
\end{equation}

If we neglect the interaction between isobars and nucleons and between isobars themselves
on the left-hand side of the eq. (\ref{eq:3.6}) and take into account (\ref{eq:3.5}),
then the wave function
$ \psi^{\Delta N}_{[\beta_{i}\beta_{j}]}$ of $\Delta N$ system satisfies the equation
$$
\sqrt{\frac{2}{A(A-1)}}\,\left(\phi_{\mathnormal{\Delta}},(T_{1}+T_{2}+M_{1}- M_{2}-
E_{\beta}-E_{({\beta _i \beta _j})^{-1}})\,\phi_{\mathnormal{\Delta}}\right)\,
\psi^{\Delta N}_{[\beta_{i}\beta_{j}]}(x_{1},x_{2})=
$$
$$
-\int d(x_{3},...,x_{A})\,\psi^{N*}_{(\beta_{i}\beta_{j})^{-1}}(x_{3},...,x_{A})\,
\left(\phi_{\mathnormal{\Delta}},V \phi_{N}\right)\,\psi _{\beta}^N(x_{1},...,x_{A}),
$$
where $T_{1}$, $T_{2}$ and $M_{1}$, $M_{2}$  are the kinetic energy operators and masses of $\Delta$-isobar and nucleon;
$\left(\phi_{\mathnormal{\Delta}},V \phi_{N}\right)$ is the transition potential.

For this equation the analytic form of the wave function
$\psi^{\Delta N}_{[\beta_{i}\beta_{j}]}$ of the $\Delta N$ system
in the nuclei with closed shells derived for the oscillator
shell model of the nuclei with {\it ls}-coupling and one boson exchange
transition potential was given in \cite{18}.  The wave function
 $ \psi^{\Delta N}_{[\beta_{i}\beta_{j}]}$ for the
shell model with {\it jj}-coupling and the
same transition potential was obtained in the work \cite{19}.

 \section{One-particle density matrix}

According to the form of the wave function (\ref{eq:3.2}), the density matrix  may be written as
\begin{equation}
\label{eq:4.1}
\rho = \rho _{NN} + \rho _{\Delta \Delta} + \rho _{N\Delta} + \rho _{\Delta N}.
\end{equation}

We shall analyze only diagonal components of the density matrix
$\rho _{NN}$ and $\rho _{\Delta \Delta} $. Because of the orthogonality of
one-particle states, the contribution
 from the non-diagonal  components of the density matrix to the square of the modulus of
 the transition  amplitude is expected to  be small or zero.

We shall consider the first term of one-particle  density matrix  ($\ref{eq:4.1}$)
$$
\rho _{NN}(X_{1};\widetilde{X}_{1})=\int d(X_{2},...,X_{A})\Psi _{N}(X_{1},...,X_{A})
\Psi _{N}^{*}(\widetilde{X}_{1},X_{2},...,X_{A}).
$$
The wave function $\Psi _{N}\left( {X_{1} ,X_{2} ,...,X_{A}} \right)$ can be written as follows
$$
\Psi _{N}(X_{1},X_{2},...,X_{A})=\sqrt{N_{N}}\,A^{S}_{1;2...A}\,\Psi_{\beta _{1}}(X_{1})
\Psi_{\beta^{-1}_{1}}^{N}(X_{2},...,X_{A}),
$$
where the wave functions
\begin{equation}
\label{eq:4.2}
\Psi_{\beta_{i}}(X_{1})=\phi_{N}(m_{1})\psi_{\beta_{i}}(x_{1})
\end{equation}
and $\Psi _{\beta _{1}^{ - 1}} ^{N} \left( {X_{2} ,...,X_{A}} \right)$ are normalized by 1.
As a result we shall get
\begin{equation}
\label{eq:4.21}
\rho _{NN}(X_{1},\widetilde{X}{_{1}})=\phi_{N}(m_{1})\, \left [ \frac{N_{N}}{A} \
\rho _{NN}(x_{1},\widetilde{x}{_{1}})\right ]\, \phi _{N}^{*}(\widetilde{m}{_{1}}),
\end{equation}
where
$$
\rho _{NN}(x_{1},\widetilde{x}{_{1}})=
\sum_{i=1}^{A}\psi_{\beta_{i}}(x_{1})\psi^{*}_{\beta_{i}}(\widetilde{x}{_{1}}),
$$

The one-particle density matrix is used in the expression for the square of the modulus
of the direct amplitude $T_{d}$,
which has the following structure: the operator \textit{t}$_{\gamma\pi} $
acts on the particle with the coordinate $ X_{1}$, which moves over to the free nucleon state,
and the system of the particles with numbers $2,... , \textit{A}$ is a ''spectator''.
In eq. (\ref{eq:4.21}) particle ''1'' is a nucleon. Therefore, the summand
$\rho _{NN}$ corresponds to the quasifree mechanism of the reaction, which is illustrated
by the diagram in Fig. 1{\it a}. The pion production occurs at interaction of the photon
with the nucleon of the nucleus
as a result of the process $\gamma N\to N'\pi $. A spectator is a system of {\it A}--1 baryons,
forming the residual nucleus, when all particles are the nucleons.

The second summand of  one-particle  density matrix ($\ref{eq:4.1}$) is
\begin{equation}
\label{eq:4.3}
\rho _{\Delta \Delta} ( {X_{1} ;\tilde {X}_{1}} ) = \int
{d( {X_{2} ,...,X_{A}} )}\Psi _{\Delta} ( {X_{1},X_{2}
,...,X_{A}}  )\Psi _{\Delta}^{\ast} ( {\tilde {X}_{1},X_{2},...,X_{A}}).
\end{equation}

Substituting in ($\ref{eq:4.3}$) the expression ($\ref{eq:3.5}$) for the wave function $\Psi _{\Delta} $,
we shall get
$$
\rho _{\Delta \Delta} ( {X_{1} ;\tilde {X}_{1}} )=\frac{2}{A(A-1)}
\sum_{ij}\sum_{i'j'}\int d(X_{2},...,X_{A})\ \times
$$
$$
\left[(A-1)\Psi^{\Delta N}_{[\beta_{i}\beta_{j}]}(X_{1},X_{2})
\Psi^{N}_{(\beta_{i}\beta_{j})^{-1}}(X_{3},X_{4},...,X_{A})\right.\ \times
$$
$$
\times\ \Psi^{\Delta N^{*}}_{[\beta_{i'}\beta_{j'}]}(\widetilde{X}_{1},X_{2})
\Psi^{N^{*}}_{(\beta_{i'}\beta_{j'})^{-1}}(X_{3},X_{4},...,X_{A})\  \ +
$$
$$
\frac{(A-1)(A-2)}{2}\Psi^{\Delta N}_{[\beta_{i}\beta_{j}]}(X_{3},X_{2})
\Psi^{N}_{(\beta_{i}\beta_{j})^{-1}}(X_{1},X_{4},...,X_{A})\ \times
$$
\begin{equation}\label{eq:4.4}
\times\ \Psi^{\Delta N^{*}}_{[\beta_{i'}\beta_{j'}]}(X_{3},X_{2})
\left.\Psi^{N^{*}}_{(\beta_{i'}\beta_{j'})^{-1}}(\widetilde{X}_{1},X_{4},...,X_{A})\right].
\end{equation}
\begin {figure}[t]
\unitlength = 1cm
\centering
\includegraphics [width = 8cm , keepaspectratio] {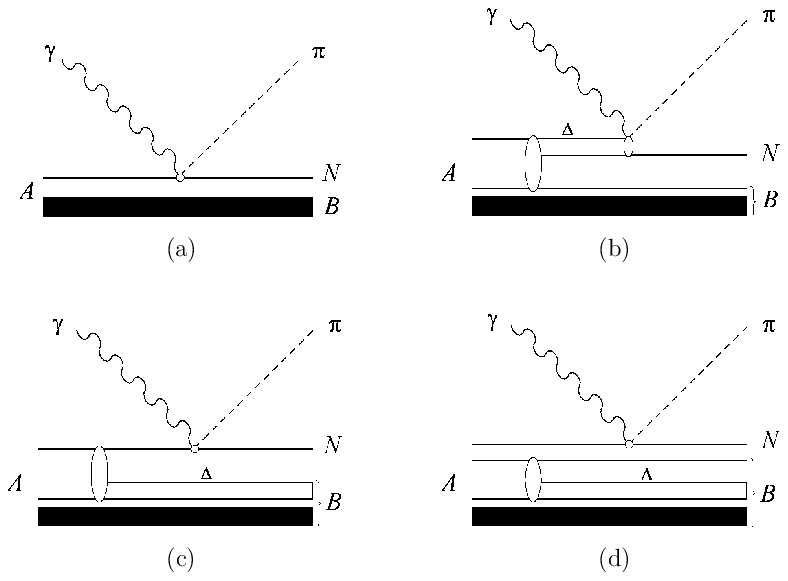}
\vspace * {7mm}
\caption { The diagrams illustrating the direct mechanisms of the $\pi N$ pair
photoproduction on nuclei in the $A(\gamma,\pi N$)B reaction. }
\label{fig1}
\end {figure}

The first summand of the formula ($\ref{eq:4.4}$) corresponds to the interaction of the photon
with $\Delta N$ system. We shall mark it $\rho ^{S}_{\Delta \Delta} ( {X_{1} ;\tilde {X}_{1}} )$.
The second  summand is $\rho ^{C}_{\Delta \Delta} ( {X_{1} ;\tilde {X}_{1}})$,
corresponds to the interaction of the photon with  the nucleon core.
As there is orthogonality of one-particle wave function, we shall get
the expression for $\rho ^{S}_{\Delta \Delta} ( X_{1} ;\tilde {X}_{1}) $
after the integration over the variable of the particles with numbers $3,...,A$
$$
\rho ^{S}_{\Delta \Delta} ( {X_{1} ;\tilde {X}_{1}} ) =
\frac{{2}}{{A}}_{} \sum\limits_{ij}^{} {\int {dX_{2}}  ^{}\Psi _{[
{\beta _{i} \beta _{j}}  ]}^{\Delta N} ( {X_{1} ,X_{2}}
)^{}\Psi _{[ {\beta _{i} \beta _{j}} ]}^{\Delta N\ast}
( {\tilde {X}_{1} ,X_{2}} )} .
$$

We will write the expression for  the second summand
in ($\ref{eq:4.4}$) $\rho ^{C}_{\Delta \Delta}( {X_{1} ;\tilde {X}_{1}})$ as
$$
\rho ^{C}_{\Delta \Delta}  ( {X_{1} ;\tilde {X}_{1}}) =
\frac{{A - 2}}{{A}} \sum\limits_{ij} {N_{\Delta\, ij} ^{}\int
{d( {X_{4} ,...,X_{A}})} ^{}\Psi _{( {\beta _{i} \beta
_{j}} )^{ - 1}}^{N} ( {X_{1} ,X_{4} ,...,X_{A}} )^{} }\ \times
$$
$$
\times \ {\Psi _{( {\beta _{i} \beta _{j}} )^{ - 1}}^{N\ast}  ( {\tilde
{X}_{1} ,X_{4} ,...,X_{A}} } ),
$$
where $N_{\Delta\, ij} $ is the norm of  wave function
$\Psi _{[ {\beta _{i} \beta _{j}} ]}^{\Delta N} $, satisfying the relationship
$$
\sum\limits_{ij} {N_{\Delta ^{}ij} = N_{\Delta} }  .
$$

 Writing the wave function of the nucleon core
 $\Psi _{\left( {\beta _{i} \beta _{j}} \right)^{ - 1}}^{N}$ in the form of decomposition
$$
\Psi^{N} _{( {\beta _{i} \beta _{j}}  )^{ - 1}} ( X_{1}
,X_{4} ,...,X_{A} ) = A_{1;2...( {A - 1} ) \ne ij}^{S}
\Psi _{\beta _{1}} (X_{1})\Psi^{N} _{( {\beta
_{i} \beta _{j} \beta _{1}}  )^{ - 1}} ( {X_{4} ,...,X_{A}})
$$
and performing integration  over the variables $X_{4} ,...,X_{A}$, we will get
$$
\rho ^{C}_{\Delta \Delta} ( {X_{1} ;\tilde {X}_{1}} ) =
\frac{{1}}{{A}}_{} \sum\limits_{ij,k \ne ij}^{} {N_{\Delta ^{}ij} ^{}\Psi
_{\beta _{k}} ^{} \left( {X_{1}}  \right)^{}\Psi _{\beta _{k}} ^{\ast}
( {\tilde {X}_{1}} )} .
$$

Using ($\ref{eq:3.5.1}$) and ($\ref{eq:4.2}$), we will write
the summands of the expressions ($\ref{eq:4.4}$) as follows
$$
\rho _{\Delta \Delta} = \rho ^{\Delta} _{\Delta \Delta}  + \rho ^{N}_{\Delta \Delta}
 + \rho ^{C}_{\Delta \Delta},
$$
where
$$
\rho ^{\Delta} _{\Delta \Delta} ( {X_{1} ;\tilde {X}_{1}} ) =
\phi _{\Delta}( {m_{1}} )\left[ { {
\,\frac{{1}}{{A}}
\int {dx_{2}\,}
\rho^{\Delta N}(x_1,x_2;\tilde {x}_{1}, {x}_{2})
}}\right]\phi _{\Delta} ^{\ast}  ( {\tilde {m}_{1}} ),
$$
$$
\rho ^{N} _{\Delta \Delta}( {X_{1} ;\tilde {X}_{1}}) =
\phi _{N}  ( {m_{1}} )\left[ {{
\,\frac{{1}}{{A}}
\int {dx_{2}\,}
\rho^{\Delta N}(x_2,x_1;{x}_{2}, \tilde {x}_{1})
}} \right]\phi _{N} ^{\ast}  ( {\tilde {m}_{1}} ),
$$
$$
\rho ^{C} _{\Delta \Delta} ( {X_{1} ;\tilde {X}_{1}} ) =
\phi _{N}  ( m_{1} )\left[
\,\frac{{1}}{{A}}
\sum\limits_{ij} {N_{\Delta ij} \
\rho^{N}_{(\beta _i \beta _j)^{-1}}(x_{1},\widetilde{x}{_{1}})
} \right]\phi _{N} ^{\ast} ( {\tilde {m}_{1}} ),
$$
$$
\rho^{\Delta N}(x_1,x_2;\tilde {x}_{1},\tilde {x}_{2})=
\sum\limits_{ij} \rho^{\Delta N}_{[ij]}(x_1,x_2;\tilde {x}_{1},\tilde {x}_{2}),
$$
$$
\rho^{\Delta N}_{[ij]}(x_1,x_2;\tilde {x}_{1},\tilde {x}_{2})= {
{  \psi _{\left[ {\beta _{i} \beta _{j}}
\right]}^{\Delta N} ( {x_{1} ,x_{2}} )^{}\psi _{\left[ {\beta
_{i} \beta _{j}}  \right]}^{\Delta N\ast} ({ \tilde {x}_{1},\tilde {x}_{2}}
)}},
$$
$$
\rho^{N}_{(\beta _i \beta _j)^{-1}}(x_{1},\widetilde{x}{_{1}})=
\sum_{k\neq ij}^{A}\psi_{\beta_{k}}(x_{1})\psi^{*}_{\beta_{k}}(\widetilde{x}{_{1}}).
$$

In the expression for the density matrix   $\rho ^{\Delta} _{\Delta \Delta}( {X_{1} ;
\tilde {X}_{1}})$ the particle ''1'' is an isobar. The reaction mechanism
corresponding to the $\rho ^{\Delta} _{\Delta \Delta} ( {X_{1} ;\tilde {X}_{1}} )$,
is illustrated by the diagram in Fig. \ref{fig1}{\it b}. In this case, the production of the pion results
from the process $\gamma \Delta \to N\pi $, under which the virtual isobar taking up photon
moves over to the real state and decays on the nucleon and the pion.

In expressions for the density matrix   $\rho ^{N} _{\Delta \Delta}( {X_{1} ;
\tilde {X}_{1}} )$ and $\rho ^{C}_{\Delta \Delta} ( {X_{1} ;\tilde {X}_{1}}
)$ the particle ''1'' is nucleon. The reaction mechanisms
corresponding to the $\rho ^{N} _{\Delta \Delta} ( {X_{1} ;
\tilde {X}_{1}} )$ and $\rho ^{C}_{\Delta \Delta} ( {X_{1} ;\tilde {X}_{1}}
)$ condition of
the isobar configurations are illustrated by the diagrams in Fig. \ref{fig1}{\it c} and  Fig. \ref{fig1}{\it d}.
They differ by the composition and the condition of the baryons, forming the remaining nuclei.

\section{Two-particle density matrix}

Two-particle density matrix is used in the expressions for the square of the modulus
 of the exchange amplitude (\ref{eq:2.5}) for the reaction $A( {\gamma ,^{}\pi N})B$.
For calculation of the density matrix
\begin{equation}\label{eq:5.1}
\rho _{NN}( {X_{1} ,X_{2} ;\tilde {X}_{1} ,\tilde {X}_{2}} ) =
\int {d( {X_{3},...,X_{A}})} ^{}\Psi _{N}( {X_{1} ,X_{2}
,...,X_{A}})^{}\Psi _{N} ^\ast ( {\tilde {X}_{1} ,\tilde
{X}_{2},...,X_{A}} )
\end{equation}
we will present the wave function $\Psi _{N}$  in the form of the expansion
$$
\Psi _{N}(X_{1} ,X_{2},X_{3} ,...,X_{A})=\left(\frac{2N_{N}}{A(A-1)}\right)^{1/2}\sum_{ij}(-1)^{i+j+1}
\Psi^{N}_{\beta_{i}\beta_{j}}(X_{1},X_{2})\ \times
$$
\begin{equation}\label{eq:5.2}
\times \ \Psi^{N}_{(\beta_{i}\beta_{j})^{-1}}(X_{3} ,...,X_{A}),
\end{equation}
where
$$
\Psi _{\beta _{i} \beta _{j}} ^{N} ( {X_{1} ,X_{2}} ) = A_{1;2}
 \Psi _{\beta _{i}} ( {X_{1}}) \Psi _{\beta _{j}}( X_{2} ).
$$

After substituting (\ref{eq:5.2}) in (\ref{eq:5.1}), taking into account
(\ref{eq:4.2}), we will get
$$
\rho _{NN}( {X_{1} ,X_{2} ;\tilde {X}_{1} ,\tilde {X}_{2}} ) =
$$
$$
\phi _{NN} \left( {m_{1} ,m_{2}}  \right)\left[ {\frac{{2N_{N}} }{{A\left(
{A - 1} \right)}}_{} \sum\limits_{ij} {\psi _{\beta _{i} \beta _{j}} ^{N}
\left( {x_{1} ,x_{2}}  \right)^{}} \psi _{\beta _{i} \beta _{j}} ^{N\ast}
\left( {\tilde {x}_{1} ,\tilde {x}_{2}}  \right)} \right]\phi _{NN}^{\ast}
\left( {\tilde {m}_{1} ,\tilde {m}_{2}}  \right).
$$

We shall go to the consideration of the density matrix   $\rho _{\Delta \Delta} $
$$
\rho _{\Delta \Delta} ( {X_{1} ,X_{2} ;\tilde {X}_{1} ,\tilde {X}_{2}}
) = \int {d( {X_{3} ,...,X_{A}})} ^{}\Psi _{\Delta
}( {X_{1} ,X_{2} ,...,X_{A}} )^{}\Psi _{\Delta}^ \ast{
}( {\tilde {X}_{1} ,\tilde {X}_{2},...,X_{A}} ).
$$

As a result of transformations of this expression, executed similarly in the previous section,
the density matrix $\rho _{\Delta \Delta} $ may be  written  as
$$
\rho _{\Delta \Delta}  = \rho ^{\Delta N}_{\Delta \Delta}  + \rho ^{N\Delta
}_{\Delta \Delta}  + \rho ^{\Delta C}_{\Delta \Delta}  + \rho ^{NC}_{\Delta
\Delta}  + \rho ^{CN}_{\Delta \Delta}  + \rho ^{C\Delta} _{\Delta \Delta}  +
\rho ^{CC}_{\Delta \Delta}  ,
$$
where
$$
\rho ^{\Delta N} _{\Delta \Delta}( {X_{1} ,X_{2} ;\tilde {X}_{1},\tilde {X}_{2}} ) =
\phi _{\Delta N} ( {m_{1} ,m_{2}})
\left[ {
\frac{{1}}{{A\left( {A - 1} \right)}}\
\rho^{\Delta N}(x_1,x_2;\tilde {x}_{1},\tilde {x}_{2})
}\right]
\phi _{\Delta N}^{\ast}  ( {\tilde {m}_{1} ,\tilde {m}_{2}} ),
$$
$$
\rho ^{N\Delta} _{\Delta \Delta} ( {X_{1} ,X_{2} ;\tilde {X}_{1},\tilde {X}_{2}}) =
\phi _{\Delta N} ( {m_{2} ,m_{1}})
\left[ {
\frac{{1}}{{A\left( {A - 1} \right)}}\
\rho^{\Delta N}(x_2,x_1;\tilde {x}_{2},\tilde {x}_{1})
} \right]
\phi _{\Delta N}^{\ast}  \left( {\tilde {m}_{2} ,\tilde {m}_{1}}  \right),
$$
$$
\displaylines{
 \rho ^{\Delta C}_{\Delta \Delta} ( {X_{1} ,X_{2} ;\tilde {X}_{1},\tilde {X}_{2}} ) =
 \cr
 \phi _{\Delta N} ( {m_{1} ,m_{2}})
 \left[ {
 \frac{{1}}{{A({A - 1})}}
 \sum\limits_{ij} {\int {d( {x_{3}})}}
 \rho ^{\Delta N}_{[ij]}(x_{1},x_{3};\tilde {x}_{1},x_{3})\,
 \rho ^N_{(ij)^{-1}}(x_{2};\tilde {x}_{2})
 } \right]
 \phi _{\Delta N}^{\ast} ( {\tilde {m}_{1},\tilde {m}_{2}}),
 \cr}
$$
$$
\displaylines{
 \rho ^{NC}_{\Delta \Delta} ( {X_{1} ,X_{2} ;\tilde {X}_{1} ,\tilde{X}_{2}}) =
 \cr
  \phi _{NN} ( {m_{1} ,m_{2}})
 \left[ {
 \frac{{1}}{{A({A - 1})}}
 \sum\limits_{ij} {\int {d( {x_{3}})}}
 \rho ^{\Delta N}_{[ij]}(x_{3},x_{1};x_{3},\tilde {x}_{1})\,
 \rho ^N_{(ij)^{-1}}(x_{2};\tilde {x}_{2})
 } \right]
 \phi _{NN}^{\ast} ( {\tilde {m}_{1},\tilde {m}_{2}}),
\cr}
$$
$$
\displaylines{
 \rho ^{CN}_{\Delta \Delta} ( {X_{1} ,X_{2} ;\tilde {X}_{1} ,\tilde{X}_{2}} ) =
 \cr
  \phi _{NN} ( {m_{1} ,m_{2}})
 \left[ {
 \frac{{1}}{{A({A - 1})}}
 \sum\limits_{ij} {\int {d( {x_{3}})}} \,
 \rho ^{\Delta N}_{[ij]}(x_{3},x_{2};x_{3},\tilde {x}_{2})\,
 \rho ^N_{(ij)^{-1}}(x_{1};\tilde {x}_{1})
 } \right]
 \phi _{NN}^{\ast} ( {\tilde {m}_{1},\tilde {m}_{2}}),
\cr}
$$
$$
\displaylines{
 \rho ^{C\Delta}_{\Delta \Delta} ( {X_{1} ,X_{2} ;\tilde {X}_{1} ,\tilde{X}_{2}} ) =
  \cr
  \phi _{\Delta N} ( {m_{2} ,m_{1}})
 \left[ {
 \frac{{1}}{{A({A - 1})}}
 \sum\limits_{ij} {\int {d( {x_{3}})}} \,
 \rho ^{\Delta N}_{[ij]}(x_{2},x_{3};\tilde {x}_{2},{x}_{3})\,
 \rho ^N_{(ij)^{-1}}(x_{1};\tilde {x}_{1})
 } \right]
 \phi _{\Delta N}^{\ast} ( {\tilde {m}_{2},\tilde {m}_{1}}),
\cr}
$$
$$
\rho ^{CC}_{\Delta \Delta}( {X_{1} ,X_{2} ;\tilde {X}_{1} ,\tilde{X}_{2}}) =
$$
$$
\phi _{NN} ( {m_{1} ,m_{2}})
\left[{
\frac{{2}}{{A\left( {A - 1} \right)}}
\sum\limits_{ij} {
N_{\Delta ij}\
\rho ^{NN}_{(ij)^{-1}}(x_{1},x_{2};\tilde {x}_{1} ,\tilde {x}_{2})
 } } \right]
 \phi _{NN}^{\ast}  ({\tilde {m}_{1} ,\tilde {m}_{2}} ),
$$
$$
\rho ^{NN}_{(ij)^{-1}}(x_{1},x_{2};\tilde {x}_{1} ,\tilde {x}_{2}) =
\sum\limits_{kl \ne ij}
{\psi _{\beta _{k} \beta _{l}} ^{N} ( {x_{1},x_{2}})}
\psi _{\beta _{k} \beta _{l}} ^{N\ast} ({\tilde {x}_{1} ,\tilde {x}_{2}} ).
$$

\begin {figure}[h]
\unitlength = 1cm
\centering
\includegraphics [width = 8cm , keepaspectratio] {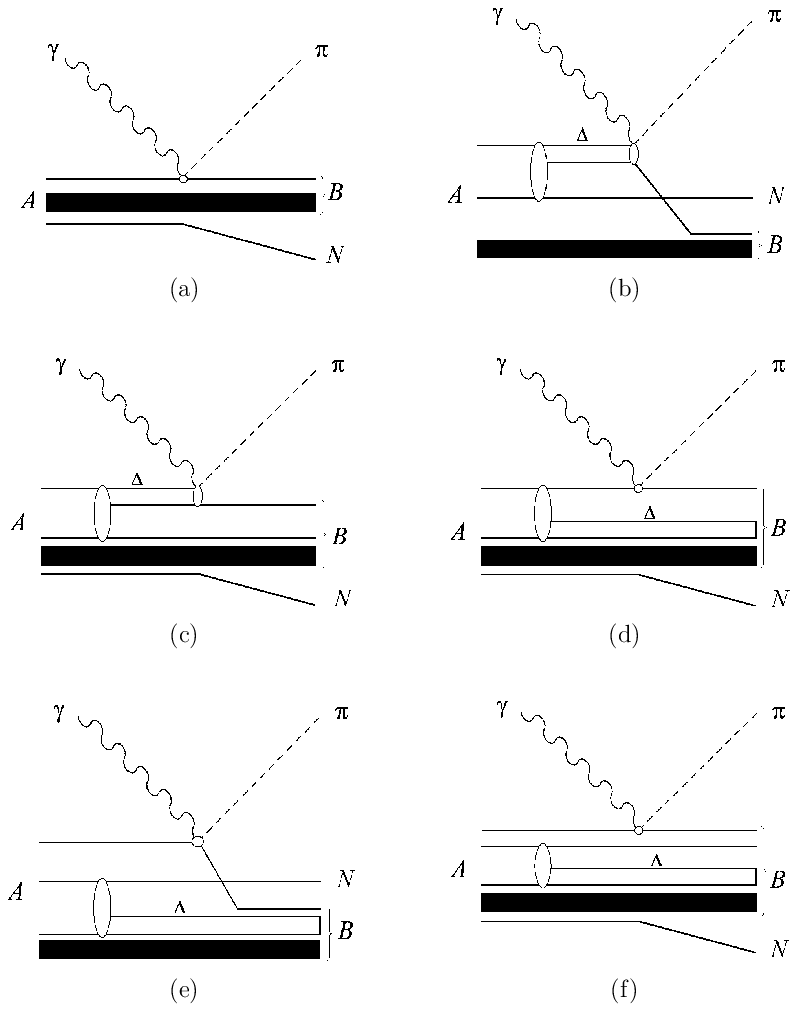}
\vspace * {4mm}
\caption { The diagrams illustrating the exchange mechanisms of the $\pi N$ pair
photoproduction on nuclei in the $A(\gamma,\pi N$)B reaction.}
\label{fig2}
\end {figure}

The exchange  amplitudes, corresponding to the matrix $\rho ^{N\Delta} _{\Delta \Delta} $
and $\rho ^{C\Delta} _{\Delta \Delta} $ are  zero, because of the orthogonality of the
wave functions $\phi _{N} \left( {m_{2}} \right)$ and $\phi _{\Delta} \left( {m_{2}} \right)$.
The remaining six summands $\rho _{NN},$ $ \rho ^{\Delta N}_{\Delta \Delta},$ $
\rho ^{\Delta C}_{\Delta \Delta},$ $ \rho ^{NC}_{\Delta \Delta},$ $\rho ^{CN}_{\Delta \Delta} $
and $\rho ^{CC}_{\Delta \Delta} $ of the two-particle density matrix  correspond
to the mechanism of the  reactions in the usual space, which  are illustrated by the diagram
shown in Fig. \ref{fig2} in the same order.

\section{Transition operator }

With the help of the formula
$$
<x'|t_{\gamma B \pi}|x>=\sum_{m'm}\phi^{*}_{N}(m')<X'|t_{\gamma \pi}|X>\phi_{B}(m),
$$
matrix element of the operator $t_{\gamma \pi}$ between
the one-particle intrinsic states defines the transition operator  $<x'|t_{\gamma B \pi}|x>$
in the configuration space, which acts on the space $\mathbf{r}$, spin {\it s}, and isospin {\it t} coordinates.

Using the {\it S}-matrix approach for the description of an elementary process $\gamma B\rightarrow N\pi$,
we will find the transition operator $t_{\gamma B \pi} $. We will suppose that {\it S}-matrix has the standard
expansion in power of the interaction Lagrangian, in which the strong interaction fields of the nucleon, pion and
isobar, the vector potential of photon are the operators acting on the space $\mathbf{r}$, spin {\it s}, and
isospin {\it t} coordinates. We will write the Lagrangian of the strong interaction baryon fields  and the pion field as
$ L_{s}=\textbf{j}_{\pi}(\textbf{r},t)\boldsymbol{\phi}(\textbf{r},t)$, where
$\textbf{j}_{\pi}(\textbf{r},t)$ is the pion current. The Lagrangian of the electromagnetic  interaction
may be written as $ L_{\gamma}=j_{\mu}(\textbf{r},t)A^{\mu}(\textbf{r},t)$, where
$ j_{\mu}(\textbf{r},t)$ is the electromagnetic current. Then, the transition operator
$t_{\gamma B \pi}$ may be written as
$$
t_{\gamma B \pi}= \boldsymbol{\varphi}_{a}^{+}
\textbf{t}_{\gamma B \pi}^{\mu}\
{\epsilon}^{\lambda}_{\mu}.
$$
Here  ${\epsilon}^{\lambda}$ is  4-vector of the photon polarization;
$\boldsymbol{\varphi_{a}}$ is the covariant unit vector of the cyclical basis describing
the isotopic state of the pion; index {\it a} takes on the values +, 0, -, which fit with the
positive, neutral and negative pions;
\begin{equation}\label{eq:6.3}
\textbf{t}_{\gamma B \pi}^{\mu}=
\int d\textbf{r}d\textbf{r}'e^{i(\textbf{p}_{\gamma}\,\textbf{r}-\textbf{p}_{\pi}\,\textbf{r}')}\int dt
e^{-iE_{\gamma}t}T(j_{\mu}(\textbf{r},t)\textbf{j}_{\pi}(\textbf{r}',0))|_{B\rightarrow N},
\end{equation}
where index $B\rightarrow N$ under {\it T}-product of the currents designates that it necessarily  leaves those
 components of the electromagnetic and pion currents, which give rise to the $ N\rightarrow N$ and
the $ \Delta\rightarrow N $ transitions. The expression of the current comes from  the interaction Lagrangian
 constructing from the photon, nucleon, $\Delta$-isobar and pion fields.
Using the operator ($\ref{eq:6.3}$), we will go to the momentum representation of the matrix element
$<x'|\textbf{t}_{\gamma B \pi}^{\mu}|x>$ and insert the complete set of the intermediate single-particle
baryon and meson states with
the minimal masses of the baryons and the pions. As a result, we will obtain
$$
 <x'|\textbf{t}_{\gamma B \pi}^{\mu}|x>=\int d\textbf{p}_{B}e^{i\textbf{p}_{N}\textbf{r}'}
  \sum_{m_{\sigma_{N}},m_{\tau_{N}},m_{\sigma_{B}},m_{\tau_{B}}}\xi^{*}_{m_{\sigma_{N}},m_{\tau_{N}}}(s',t')\,\times
 $$
 \begin{equation}\label{eq:6.4}
 \times\,<m_{\sigma_{N}},m_{\tau_{N}}|\textbf{T}^{\mu}_{\gamma B\rightarrow N \pi}(\textbf{p}_{N},\textbf{p}_{B})|m_{\sigma_{B}},m_{\tau_{B}}>
 \xi_{m_{\sigma_{B}},m_{\tau_{B}}}(s,t)e^{-i\textbf{p}_{B}\textbf{r}}.
\end{equation}
Here {\it B} and {\it N} are the indices of the initial baryon and the final nucleon;
$ \textbf{p}_{B}$ and $\textbf{p}_{N}$ are the momenta of the baryon and the nucleon  for the process
$\gamma B\rightarrow \pi N$, $\textbf{p}_{N}=\textbf{p}_{B}+\textbf{q}$,
where $\textbf{q}=\textbf{p}_{\gamma}-\textbf{p}_{\pi}$ is the transfer  momentum,\,
 $ \xi _{m_{\sigma_{N}},m_{\tau_{N}}}$ and\, $\xi_{m_{\sigma_{B}},m_{\tau_{B}}} $ are
the spin-isospin wave functions of the nucleon and the baryon. If it designates:
$\alpha\equiv\textbf{p},m_{\sigma},m_{\tau}$ is the baryon state index; {\it n} is
the index of the intermediate baryon, $ \alpha_{\pi^{a}}$ is the index of the pion, then
$$
<m_{\sigma_{N}},m_{\tau_{N}}|\textbf{T}^{\mu}_{\gamma B\rightarrow N \pi}(\textbf{p}_{N},\textbf{p}_{B})|
m_{\sigma_{B}},m_{\tau_{B}}>\ =
$$
$$
\frac{(2\pi)^{3}}{i}\left[\sum_{n=N,\Delta}\frac{<\alpha_{N}|j^{\mu}(0,0)|\alpha_{n}>
<\alpha_{n}|\textbf{j}_{\pi}(0,0)|\alpha_{B}>}{E_{\gamma}+E_{n}-E_{N}-i\varepsilon}\right. +
$$
$$ +
\sum_{\widetilde{n}=N,\Delta}\frac{<\alpha_{N}|\textbf{j}_{\pi}(0,0)|\alpha_{\widetilde{n}}>
<\alpha_{\widetilde{n}}|j^{\mu}(0,0)|\alpha_{B}>}{E_{\widetilde{n}}-E_{\gamma}-E_{B}-i\varepsilon}-
$$
$$
-\left. 2\frac{E^{2}_{\widetilde{\pi}}}{E_{\pi}}\sum_{b}\frac{<\alpha_{\pi^{a}}|j^{\mu}(0,0)|\alpha_{\widetilde{\pi}^{b}}>
<\alpha_{N}|\textbf{j}_{\pi}(0,0)|\alpha_{B}>}
{E^{2}_{\widetilde{\pi}}-\textbf{p}^{2}_{\widetilde{\pi}}-m^{2}_{\pi}}+
i<\alpha_{N}|\textbf{j}^{\mu}_{\gamma\pi NB}|\alpha_{B}>\ \right],
$$
where $\textbf{p}_{n}=\textbf{p}_{B}-\textbf{p}_{\pi} $,
$\textbf{p}_{\widetilde{n}}=\textbf{p}_{\gamma}+\textbf{p}_{B}$ and $\textbf{p}_{\widetilde{\pi}}=-\textbf{q}$
are the momenta of the intermediate baryons and mesons,
$E_{n}=(\textbf{p}^{2}_{n}+m^{2}_{n} )^{1/2}$, $E_{\widetilde{n}}=
(\textbf{p}^{2}_{\widetilde{n}}+m^{2}_{\widetilde{n}})^{1/2}$,
$ E_{\widetilde{\pi}}=E_{\gamma}-E_{\pi}$.

 The matrix elements of the currents are written by means of the nonrelativistic currents as
$$
 <\alpha_{2}|j^{\mu}(0,0)|\alpha_{1}>\,=\frac{1}{2}(E_{2}E_{1})^{-1/2}\,\xi^{*}_{m_{\sigma_{2}},m_{\tau_{2}}}
 j^{\mu ,B_{2}\leftarrow B_{1}}(\textbf{p}_{2},\textbf{p}_{1})\,\xi_{m_{\sigma_{1}},m_{\tau_{1}}},
$$
$$
 <\alpha_{2}|\textbf{j}_{\pi}(0,0)|\alpha_{1}>\,=\frac{1}{2}(E_{2}E_{1})^{-1/2}\,\xi^{*}_{m_{\sigma_{2}},m_{\tau_{2}}}
 \textbf{j}_{\pi}^{B_{2}\leftarrow B_{1}}(\textbf{p}_{2},\textbf{p}_{1})\,\xi_{m_{\sigma_{1}},m_{\tau_{1}}},
$$
$$
<\alpha_{\pi^{a}}|j^{\mu}(0,0)|\alpha_{\widetilde{\pi}^{b}}>\,=\frac{1}{2}\,(E_{\pi}E_{\widetilde{\pi}})^{-1/2}
\boldsymbol{\varphi}^{+}_{a}\, j^{\mu,\pi\leftarrow \pi}
(\textbf{p}_{\pi},\textbf{p}_{\widetilde{\pi}})\,\boldsymbol{\varphi}_{b}.
$$
The explicit expressions of nonrelativistic currents are
$$
 \textbf{j}_{\gamma} ^{N \leftarrow N} \left( {\textbf{p}_{2} ,\textbf{p}_{1}}  \right) =
\frac{{e}}{{2m_{N}} }\left[ {\left( {\textbf{p}_{2} + \textbf{p}_{1}}  \right)\frac{{1 + \tau
_{3}} }{{2}} - i\left( {\textbf{p}_{2} - \textbf{p}_{1}}  \right)\times \boldsymbol{\sigma}  \left(
{\frac{1+\tau_{3}}{2}\mu_{p}+\frac{1-\tau_{3}}{2}\mu_{n}
} \right)} \right],
$$
$$
 \textbf{j}_{\gamma} ^{N \leftarrow \Delta}  \left( {\textbf{p}_{2} ,\textbf{p}_{1}}  \right) =i
\frac{{e}}{{2m_{N} } }\mu _{\gamma N\Delta}  \left[ {\left( {\textbf{p}_{2} - \textbf{p}_{1}}
\right) \times \textbf{S}^{ +} } \right] T^{ +}_{3} ,
$$
$$
 \textbf{j}_{\gamma} ^{\Delta \leftarrow \Delta}  \left( {\textbf{p}_{2} ,\textbf{p}_{1}}  \right) =
\frac{{e}}{{2m_{\Delta} } } {\left( {\textbf{p}_{2} + \textbf{p}_{1}}  \right)\frac{{1 +
\Theta _{3}} }{{2}} - i\,  \frac{e}{2m_{N}}\, \mu _{\Delta^{++}}\left[\left( {\textbf{p}_{2} -\textbf{ p}_{1}}
\right)\times \boldsymbol{\Sigma }_{\Delta}\right]\frac{{1 + \Theta _{3}} }{{4}} },
$$
$$
 \textbf{j}_{\gamma} ^{\pi N \leftarrow \Delta}  = i_{} e_{}\, \frac{{f_{\pi N\Delta}
}}{{m_{\pi} } }\,\textbf{S}^{ +} \,\textbf{T}^{ +} ,\
\ \ \textbf{j}_{\gamma} ^{\pi N \leftarrow N} = i_{}
e_{}\, \frac{{f_{\pi NN}} }{{m_{\pi} } }\,\boldsymbol{\sigma} \boldsymbol{\tau} ,
$$
$$
 \textbf{j}_{\gamma} ^{\pi \leftarrow \pi}  \left( {\textbf{p}_{2} ,\textbf{p}_{1}}  \right) = \frac{e}{2}
 \left( {\textbf{p}_{2} - \textbf{p}_{1}}  \right),\
\ \textbf{j}_{\pi} ^{\Delta \leftarrow
\Delta}  \left( {\textbf{p}_{2} ,\textbf{p}_{1}}  \right) = -i\,\frac{{f_{\pi \Delta \Delta}
}}{{m_{\pi} } }\, \left( {\textbf{p}_{2} - \textbf{p}_{1}}  \right)\cdot\boldsymbol{\Sigma} _{\Delta} \textbf{T}_{\Delta}  ,
$$
$$
 \textbf{j}_{\pi} ^{N \leftarrow N} \left( {\textbf{p}_{2} ,\textbf{p}_{1}}  \right) = -i\,\frac{{f_{\pi
NN}} }{{m_{\pi} } }\, \  \left( {\textbf{p}_{2} - \textbf{p}_{1}}  \right)\cdot \boldsymbol{\sigma}\boldsymbol{\tau},
$$
$$
\textbf{j}_{\pi} ^{N \leftarrow \Delta}  \left( {\textbf{p}_{2} ,\textbf{p}_{1}}  \right) =-
i\, \frac{{f_{\pi N\Delta} } }{{m_{\pi} } }\, \  \left( {\textbf{p}_{2} - \textbf{p}_{1}
} \right)\cdot\textbf{S}^{ +} \mathop {\textbf{T}^{ +} }\limits.
$$

Here $\textbf{S}$ is transition spin operator which converts the spin-1/2 state into the spin-3/2 state.
The matrix $\textbf{S}$ is defined as
$$
 \textbf{S}_{m_{\frac{3}{2}},\, m_{\frac{1}{2}}}=\sum_{m}
C_{\frac{1}{2},\,m_{\frac{1}{2}};\, 1,\,m}^{\frac{3}{2},\,m_{\frac{3}{2}}}\textbf{e}_{m},
$$
 where  the Condon and Shortly phase convention for Clebsch-Gordon coefficients has been used and
 $\textbf{e}_{m}$ is unit vector in the spherical basis. The matrix $\textbf{T}$ is transition
 isospin operator which converts the isospin-1/2 state into the isospin-3/2 state. The matrices
 $\boldsymbol{\Sigma }_{\Delta}$  and $\textbf{T}_{\Delta}$ are defined as
$$
\boldsymbol{\Sigma }_{\Delta}=\sum_{l=1}^{3}S_{l}\boldsymbol{\sigma }S_{l}^{+}, \ \ \ \
 \textbf{T}_{\Delta}=\sum_{l=1}^{3}T_{l}\boldsymbol{\tau }T_{l}^{+}.
$$
We shall notice that the matrices $\boldsymbol{\Sigma }_{\Delta}$  and $\textbf{T}_{\Delta}$ are
$$
\boldsymbol{\Sigma }_{\Delta}=\frac{\boldsymbol{\Sigma }}{3} \hspace{2mm} \mbox{and}\hspace{2mm}
\textbf{T}_{\Delta}=\frac{\boldsymbol{\Theta}}{3},
$$
where $\boldsymbol{\Sigma }(\boldsymbol{\Theta})$ is the Pauli operator for the $\Delta$ spin(isospin).
   The proton and neutron magnetic moments of the
nucleon  are  accordingly $\mu _{p}=2.79$ and $\mu _{n}=-1.91$ in terms of nuclear magnetons.
The magnetic moment of the isobar is taken to be $\mu_{\Delta^{++}}$= 4.52 in terms of nuclear magnetons,
using the value obtained from a soft-photon analysis of pion-proton bremsstrahlung data near the $\Delta^{++}$
resonance [20]. Using the excitation strengths $G_{M}$=0.28, as obtained from the analysis of $\gamma N$ data
in the $\Delta$-resonance region [21], we calculated  the value of the transition magnetic moment
$\mu_{\gamma N\Delta}$=3.42 in terms of nuclear magnetons.
The $\pi NN$ coupling constants is
$f_{\pi NN}^{\,2} / 4\pi  = 0.08$. For the $\pi N\Delta$ coupling constant we take the value $f_{\pi N\Delta}$  =
2.123 obtained from the decay $\Delta\rightarrow \pi N$ [22]. The $\pi \Delta\Delta$ coupling constants is
 $f_{\pi \Delta \Delta}$ = $4/5f_{\pi NN}$, as predicted by the trivial quark model.

The result ($\ref{eq:6.4}$) for $<x'|\textbf{t}_{\gamma B \pi}^{\mu}|x>$ answers, in general,
to the non-locale  interaction that brings about the need of the calculation of multivariate integral
on $\textbf{p}_{B} ,$ $ \textbf{r}',$ $ \textbf{r}$. Supposing the momentum $\textbf{p}_{N}$ in the
operator $ \textbf{T}_{\gamma B \to N\pi} ^{\mu}  ( {\textbf{p}_{N} ,\textbf{p}_{B}}
) $ is fixed $\textbf{p}_{N}=\widetilde{\textbf{p}}$ gives the
local transition operator $ < x'|t_{\gamma{\rm B}\pi} ^{\mu}  |x > $
$$
 < x'|t_{\gamma {\rm B}\pi} ^{\mu}  |x > _{} = _{} \delta ( {\textbf{r}' - \textbf{r}}
)_{} e^{i\,( {\textbf{p}_{\gamma}  \textbf{r} - \textbf{p}_{\pi}  \textbf{r}'} )} <
s',t'|T_{\gamma {\rm B}\pi} ^{\mu}  ( {\tilde {\textbf{p}},\tilde {\textbf{p}} - \textbf{q}})|s,t > .
$$
Here
$$
 < s',t'|\textbf{T}_{\gamma {\rm B}\pi} ^{\mu}  ( {\tilde {\textbf{p}},\tilde {\textbf{p}} - \textbf{q}}
)|s,t > \,=\, \sum_{m_{\sigma_{N}},m_{\tau_{N}},m_{\sigma_{B}},m_{\tau_{B}}}
{\zeta _{m_{\sigma_{N}},m_{\tau _{N}}} ^{\ast}  ({s',t'})}\, \times
$$
$$
\times \ <m_{\sigma_{N}},m_{\tau_{N}}|\textbf{T}^{\mu}_{\gamma B\rightarrow N \pi}
({\tilde {\textbf{p}},\tilde {\textbf{p}} - \textbf{q}} )|m_{\sigma_{B}},m_{\tau_{B}}>
\zeta _{m_{\sigma_{B}},m_{\tau_{B}}}  ({s,t}).
$$

Calculating the amplitude of the transition $\gamma B \to N\pi $, as momentum  $\widetilde{\textbf{p}}$,
we shall take the momentum, which is canonically conjugate to the coordinate
\textbf{r} in the nucleon wave function. For the amplitude,
corresponding to the diagram on fig. \ref{fig1}{\it a}, the momentum $\tilde {\textbf{p}}$
is the momentum $\textbf{p}_{n} $ of the free nucleon.
In the exchange amplitude  the momentum $\tilde {\textbf{p}}$ is the integration variable.

Using the explicit form of the  expression of non-relativistic currets, the transition operator
$ t_{\gamma \Delta\pi}$
may be written as
$$
t_{\gamma \Delta \pi}=\boldsymbol{\varphi}_{a}^{+}\,\sum^{4}_{i=1}\sum^{3}_{j=1}f_{ij}\,M_{i}\,\textbf{I}_{j},
$$
where $M_{i}$ are the independent spin structures
$$
\begin{array}{ll}
 M_{1}=\boldsymbol{\varepsilon}^{\lambda}\cdot\mathbf{S}^{+}; &
 M_{2}=i\boldsymbol{\sigma}\cdot[\mathbf{p}_{\gamma}\times\boldsymbol{\varepsilon}^{\lambda}]\,\mathbf{p}_{\pi}\cdot
\mathbf{S}^{+};
\\ & \\
 M_{3}=\mathbf{p}_{\pi}\cdot \boldsymbol{\varepsilon}^{\lambda}\,\mathbf{p}_{\gamma}\cdot
 \mathbf{S}^{+}; \ \ \ \ \ \ \ &
M_{4}=\mathbf{p}_{\pi}\cdot \boldsymbol{\varepsilon}^{\lambda}\,\mathbf{p}_{\pi}\cdot
 \mathbf{S}^{+}.
\end{array}
$$
Here $\boldsymbol{\varepsilon}^{\lambda}$ is  3\,-vector of the photon polarization.

The isospin structures are
$$
\mathbf{I}_{1}=\mathbf{T}^{+},\ \mathbf{I}_{2}=\boldsymbol{\tau}T^{+}_{3},\
 \mathbf{I}_{3}=\tau_{3}\mathbf{T}^{+}
$$
The values $f_{ij}$ are
$$
\begin{array}{ll}
\displaystyle f_{11}=\alpha\,\left[-\frac{\mu _{\Delta^{++}}}{6\,a}\,\mathbf{p}_{\pi}\mathbf{p}_{\gamma}+2\,m_{N}\right],
&
\displaystyle f_{12}=\alpha\left[\frac{\mu_{\Delta^{++}}}{3\,a}-
\frac{F}{3\,b}\right]\mathbf{p}_{\pi}\mathbf{p}_{\gamma},
\\
\displaystyle f_{13}=\alpha\left[-\frac{\mu_{\Delta^{++}}}{2\,a}+
\frac{2F}{9\,b} \right]\mathbf{p}_{\pi}\mathbf{p}_{\gamma},
&
\displaystyle f_{21}=\alpha\left[-\frac{\mu_{\Delta^{++}}}{12\,a}+\frac{\mu_{p}+\mu_{n}}{2\,c}\right],
\\
\displaystyle f_{22}=\alpha\left[\frac{\mu_{\Delta^{++}}}{6\,a}-
\frac{F}{b} \right],
&
\displaystyle f_{23}=\alpha\left[-\frac{\mu_{\Delta^{++}}}{4\,a}+\frac{2\,F}{3\,b}+
\frac{\mu_{p}-\mu_{n}}{2\,c}\right],
\\
\displaystyle f_{31}=\alpha\left[\frac{\mu_{\Delta^{++}}}{6\,a}-\frac{2\,m_{N}}{d}\right],
&
\displaystyle f_{32}=\alpha\left[-\frac{\mu_{\Delta^{++}}}{3\,a}+\frac{F}{3\,b} \right],
\\
\displaystyle f_{33}=\alpha\left[\frac{\mu_{\Delta^{++}}}{2\,a}-\frac{2\,F}{9\,b} \right],
&
\displaystyle f_{41}=\alpha\frac{2\,m_{N}}{d},
\\
\displaystyle f_{42}=0,
&
\displaystyle f_{43}=0.
\end{array}
$$
Here
$$
\begin{array}{ll}
\displaystyle F=\mu _{\gamma N\Delta}\, \frac{{f_{\pi\Delta\Delta}}}{f_{\pi N\Delta}}, &
\displaystyle \alpha=i\,\frac{e}{2m_{N}}\ \frac{f_{\pi N\Delta}}{m_{\pi}},
\\&\\
a=E_{\Delta}+E_{\gamma}-E_{a}+\frac{i}{2}\Gamma_{\Delta}, &
b=E_{\Delta}-E_{\pi}-E_{b}+i\varepsilon,
\\ & \\
c=E_{\Delta}-E_{\pi}-E_{c}+i\varepsilon, \ \ \ \ \ \ \ &
d=(\mathbf{p}_{N}-\mathbf{p}_{\Delta})^{2}-(E_{N}-E_{\Delta})^{2}+m^{2}_{\pi}+i\varepsilon,
\end{array}
$$
$$
E_{a}=((\mathbf{p}_{\Delta}+ \mathbf{p}_{\gamma})^2 + m_{\Delta}^{2})^{1/2},\,
E_{b}=((\mathbf{p}_{\Delta}- \mathbf{p}_{\pi})^2 + m_{\Delta}^{2})^{1/2},\,
E_{c}=((\mathbf{p}_{\Delta}- \mathbf{p}_{\pi})^2 + m_{N}^{2})^{1/2}.
$$

We shall use the non-relativistic operator
of Blomqvist-Laget \cite{23} as the transition operator $ t_{\gamma N \pi}$ .

 \section{ Cross section of the $A( {\gamma ,^{}\pi ^{}N})B$ reaction}

\begin{figure}[h]
\centering
\includegraphics[width=6cm,keepaspectratio]{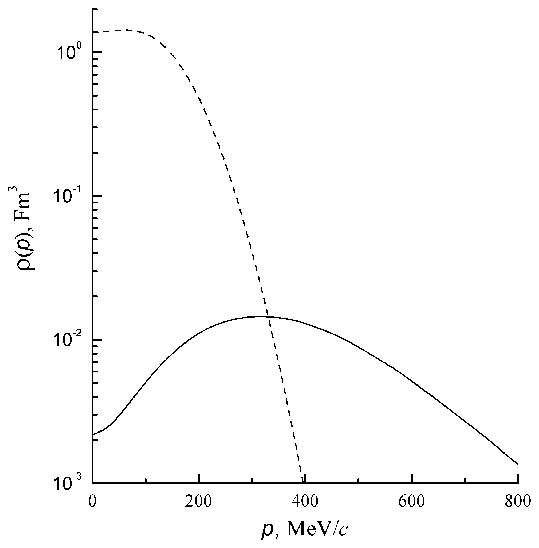}
\vspace*{8pt}
\caption{The momentum distributions of
the $\Delta$-isobar (solid curve)  and protons (dashed curve)  of nucleus  $^{12}$C}
\label{fig3}
\end{figure}

Following from the consideration of one- and two-particle density matrixes,
an account of  the isobar configurations in the atomic nuclei results in
a significant increase of  possible  direct and  exchange mechanisms of the reactions in the usual space.
One-particle  and  two-particle density matrixes
present themselves as some combinations of  one-particle wave functions of the  nucleons bound
in nuclei  $\psi _{\beta _{i}} ( {x} )$ and wave function $\psi
_{[ {\beta _{i} \beta _{j}}]}^{\Delta N} ( {x_{1} ,x_{2}})$,
describing the $\Delta N$ system. Separate components of  one-particle and two-particle
density matrixes are connected with different mechanisms of considered reaction.
For qualitative estimation of the kinematic area, where  different mechanisms
of the  reactions are shown, the momentum distributions of
the isobar $\rho ^{\Delta} $ and proton $\rho^{N}$ of nucleus  $^{12}$C defined as
$$
\rho ^{N}( {p} ) = \sum\limits_{i} {\int {dy_{_N}} \, \Phi _{\beta
_{i}} ^{\ast}  ( {y_{_N}} )_{}} \, \delta ( {p - p_{_N}}
)_{} \delta _{_{1/2},m_{\tau _i}}  \Phi _{\beta _{i}}  ({y_{_N}} )/\, p^{2},
$$
$$
\rho ^{\Delta} ( {p} ) = \sum\limits_{ij} {\int {d(y_{_\Lambda},y_{_N}) \,}
\Phi _{[ {\beta _{i} \beta _{j}} ]}^{\Delta N\ast} \, (
{y_{_\Delta} ,y_{_N}}  )_{}}  \delta ( {p - p_{_\Delta}} )_{} \Phi
_{[ {\beta _{i} \beta _{j}} ]}^{\Delta N} ( {y_{_\Delta} ,y_{_N}})/\, p^{2}
$$
are given in Fig. \ref{fig3},
where $y \equiv \mathbf{p},s,t$; $\Phi _{\beta _{i}} ( {y}_{_N} )$ and
$\Phi _{[{\beta _{i} \beta _{j}} ]}^{\Delta N} ( {y_{_\Lambda} ,y_{_{_N}}})$
are the Fourier transforms of the wave functions
$\psi _{\beta _{i}} ({x}_{_N} )$ and
$\psi _{[{\beta _{i} \beta _{j}} ]}^{\Delta N} \left( {x_{_\Lambda} ,x_{_{_N}}}\right).$

Let us consider the direct mechanisms of the reaction. First of all it is necessary to note that
all four  mechanisms of the reaction shown in Fig. \ref{fig1}({\it a}--{\it d}), give  contributions
to the cross section of the photoproduction of the
$\pi ^{ +} n,$ $\pi ^{ -} p,$ $\pi ^{0}p$ and $ \pi ^{0}n$ pairs.
The contributions
to the cross section of the photoproduction of the $\pi ^{ +} p$ and $\pi ^{ -} n$
pairs are due to the  mechanism in Fig. \ref{fig1}{\it b}. So, these two reactions are perspective
for the study of the virtual states of the isobar in a nucleus. Since the density of the momentum
distribution of the nucleon under small momentum  is  by several orders more than the density
of the momentum distribution of the ${\Delta}$-isobar, the mechanisms of the reactions
corresponding to the diagrams in Fig. \ref{fig1}{\it a} and Fig. \ref{fig1}{\it d} practically completely define
the behavior of the cross section of  reactions in this kinematic area. At large momentum
transfer to the nucleus exceeding 400 MeV/{\it c},  the mechanisms of the reactions corresponding
to the diagrams in Fig. \ref{fig1}{\it b} and in Fig. \ref{fig1}{\it c} dominate. The relative contribution of these
 diagrams is defined, basically, by the probability of the $\gamma N\rightarrow N \pi$ and
$\gamma \Delta \to N\pi $ transitions. The direct mechanism of a pion photoproduction corresponding
to the diagrams in Fig. \ref{fig1}{\it a}
and taking into account only the nucleon configurations is analysed in detail in Refs. \cite{24,25,26}.

The exchange mechanisms of the reactions  shown in Fig. \ref{fig2}({\it a}--{\it f}) may be divided into  two groups.
One group includes the mechanisms in which the nucleon belonging to the
nucleon core of the nucleus becomes free.
Manifestations of the exchange mechanism of  the neutral pion photoproduction
within the framework of the model, taking into account only the nucleon configurations of the nuclei,
were analysed in the works \cite{27,28}. The contribution of the appropriate exchange  transition amplitudes
quickly decreases with growing of the nucleon energy and concentrates near the kinematic
area of the coherent pion photoproduction on the residual nucleus, in range of small momentum transfer
$\textbf{p}_{\gamma} - \textbf{p}_{\pi}$. Another group contains mechanisms of the reactions in
which the nucleon of the $\Delta N$ system becomes free.
These mechanisms of the $\pi ^{ +} p$ pair
photoproduction were considered in the work \cite{29}. The contribution of  them to the cross
section of the reactions concentrated  in the considerably  greater range of the nucleon  momentum,
practically coinciding with  range of definition of the virtual isobar momentum distribution.

The cross section  corresponding to the different pion-nucleon pair production mechanisms
is numerically estimated  in respect of $^{12}$C$( {\gamma ,\pi ^{-} p})^{11}$C and
$^{12}$C$( {\gamma ,^{}\pi ^{ +} p})^{11}$Be reactions. The differential cross
sections are calculated for that mechanisms in which the proton momentum in the final states can be
large enough  for the experimental checking of the model predictions by means of   simultaneous
registration of the pion and the proton in the experiment. These are, first of all, the direct mechanisms
of the reactions and the exchange mechanisms in which the nucleon of the $\Delta N$ systems
becomes free.

The cross section of the reaction  $^{12}$C$( {\gamma ,^{}\pi ^{ -} p})^{11}$C is calculated
 in the kinematic  area in which the experimental data \cite{10}  were interpreted \cite{11} as the manifestation of
 a quasibound isobar-nucleus states.

\begin{figure}[t]
\centering
\includegraphics [width = 8cm , keepaspectratio] {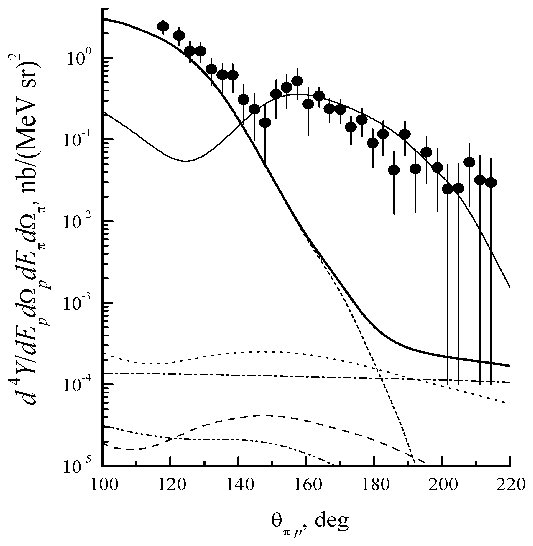}
\vspace*{-5pt}
\caption{Differential yield of the $^{12}$C$( {\gamma ,\pi ^{-} p})$ reaction for $E_{\mbox{max}}= 500$\,MeV versus the
opening angle $\theta _{\pi p}$. Thin solid curve: the $\Delta$-nucleus  model \cite{11}, in which  the existence of
the  $^{11}$B$_{\Delta} $ quasibound isobar-nucleus states is assumed in the intermediate state of the
$^{12}$C$( {\gamma ,\pi ^{-} p})p\, ^{10}$B reaction, short dashed line:  quasifree pion photoproduction,
dashed and doted curves: direct mechanisms of the $^{12}$C$( {\gamma ,\pi ^{-} p})^{11}$C reaction,
corresponding to diagrams in
Fig. \ref{fig1}\textit{b} and Fig. \ref{fig1}\textit{c}, dashed–dotted curves with one and two points:
the exchange mechanisms of the reaction, corresponding
to the diagrams in Fig. \ref{fig2}\textit{e} and Fig. \ref{fig2}\textit{b}, thick solid curve:
the total contribution to the cross section of the nucleon and isobar configurations, data are taken from Ref. \cite{10}.}
\label{fig4}
\end{figure}

The experiment \cite{10} was performed in the $\Delta(1232)$ resonance energy region using the  bremsstrahlung
photon beam from the Tomsk synchrotron. The pions and protons were detected in coincidence by two spectrometers
placed on opposite sides of the photon beam.
Figure \ref{fig4} shows the dependence of the differential reaction
yield $d^{4}Y/dE_{p} d\Omega _{p} dE_{\pi}  d\Omega _{\pi}  $ as the function of the opening angle
$\theta _{\pi p} = \theta _{\pi}  + \theta _{p} $, where $\theta _{\pi}$ and $\theta _{p}$ are
the pion and proton polar angles, connected  with differential cross section
 $$
\frac{{d^{3}\sigma} }{{dE_{p} d\Omega _{p} d\Omega _{\pi} } } = \left( {2\pi
} \right)^{ - 5}\frac{{p_{\pi} ^{3} p_{p} E_{p} E_{R}} }{{4E_{\gamma}
\left| {E_{R} p_{\pi} ^{2} - E_{\pi _{}}  p_{\pi _{}}  p_{R}}
\right|}}\overline {\left| {{\rm{T}}_{fi}}  \right|^{2}}
$$
by the relation
$$
\frac{d^{4}Y}{dE_{p} d\Omega _{p} dE_{\pi}  d\Omega _{\pi} } =
\frac{{d^{3}\sigma} }{{dE_{p} d\Omega _{p} d\Omega _{\pi} } }f(E_{\gamma})
\left|\frac{\partial E_{\gamma}}{\partial E_{\pi}}\right|.
$$
Here $f( {E_{\gamma} })$ is the bremsstrahlung spectrum,
normalized as
$$
\int {dE_{\gamma }  f\left( {E_{\gamma} }  \right)} _{} E_{\gamma}  =
E_{\mbox{max}} ,
$$
where $E_{\mbox{max}} $ is the maximal energy of the bremsstrahlung,
 $\overline {\left|{{\rm{T}}_{fi}}  \right|^{2}} $
is the square of the modulus of the reaction amplitude ${\rm{T}}_{fi}$, averaged over
photon polarization states and summed over proton and residual nucleus  states and which is connected with
the matrix element of the $T_{fi} $ in (\ref{eq:2.1}) by the relation
$$
T_{fi} = \left( {2\pi}  \right)^{3}\delta \left( {\textbf{p}_{\gamma}  - \textbf{p}_{\pi}  -
\textbf{p}_{p} - \textbf{p}_{R}}  \right){\rm{T}}_{fi} .
$$
In Fig. \ref{fig4} both the experimental data and the calculated reaction yields are averaged   over the proton
energy in the interval of $60 - 140$ MeV and over the pion energy in the interval of $102 - 148$ MeV.

Kinematically, in Fig. 4 the small opening angle region corresponds to smaller momentum transfers to the
residual nuclear system. In this kinematical region the differential yield of the quasifree photoproducton
of the pions is dominant as it is indicated by short dashed line in Fig. 4. The differential yield includes
the contributions from two mechanisms of the reaction corresponding to the diagrams in Fig. 1a and Fig. 1d.
The pion production in this case occurs at interaction of the photon with the nucleon in the state which
is lower than the Fermi level. The wave function of the nucleon bound state is calculated using the
harmonic-oscillator shell model which reproduces the charge radius of the $^{12}$C nucleus.
Final-state interaction is taken into account through optical model. As can be seen,
the agreement between the data [10] and the quasifree pion photoproduction model both for shape and magnitude
is reasonable for opening angles up to $145^{\circ}$.

In this paper we are  primarily interested in the large opening angle region, where the momentum transfer is
relatively large and where  discrepancy is observed between data \cite{10} and the quasifree
pion photoproduction model.

In Fig. \ref{fig4} the thin solid curve indicates the angular dependence of the differential reaction
yield calculated within the framework of the $\Delta$-nucleus model \cite{11}, in which  the existence
of the $^{11}$B$_{\Delta}$ quasibound isobar-nucleus states  is assumed  in the intermediate state of
the $^{12}$C$({\gamma ,\pi ^{-} p})p\, ^{10}$B reaction. The $\Delta$-nucleus model \cite{11} predicts
the functional dependence of the cross section but not its absolute value; therefore, the reaction
yield represented by the thin solid curve was normalized by fitting to  the data points in the opening
angle range $\theta _{\pi p}=145^{\circ}-215^{\circ}$.

The dashed and dotted curves present the contributions to the cross section of the  two direct mechanisms
of the reactions, corresponding to diagrams in  Fig. \ref{fig1}\textit{b} and Fig. \ref{fig1}\textit{c},
in which the product of the pions results from the interaction of the photon with the $\Delta N$ system.
The shapes of two these curves follow substantially the momentum distribution of the isobar and the nucleon
of the $\Delta N$ system.

The contributions of the exchange mechanisms of the reaction to the cross section, corresponding
to the diagrams in Fig. \ref{fig2}\textit{e} and Fig. \ref{fig2}\textit{b} are presented by the
dashed-dotted curves with one and two points.
In the isobar exchange amplitude (Fig. \ref{fig2}\textit{b}) the large opening angle region corresponds to
higher momentum transfers $\textbf{q}$ in the
$\gamma \Delta \rightarrow N \pi$ process. Since nucleon {\it N} is in bound state, the cross section decreases
quickly as the opening angle is increased. Such intercoupling of the opening angle and  momentum transfer
$\textbf{q}$ is absent in the nucleon exchange amplitude, corresponding to diagram in  Fig. \ref{fig2}\textit{e}.
The contribution of this reaction mechanism to the cross section does not depend practically on the opening angle
in the considered kinematical region.

The thick solid curve of the Fig. \ref{fig4} shows the total contribution to the cross
section of the nucleon and isobar configurations calculated with the presented approach.
It is seen that the relative contribution of the isobar configurations increases with the opening angle increase.
When the opening angle reaches $180^{\circ}$ the contribution of the isobar configurations becomes significant,
approximately  equal to the contribution of the quasifree pion photoproduction. In the opening angle range of
$\theta _{\pi p}>180^{\circ}$ the dominant contributor is the exchange mechanism of the reaction, corresponding
to the diagram in  Fig. \ref{fig2}\textit{e}.

From Fig. \ref{fig4} we see that the angular dependence shapes of the cross sections
from some mechanisms of the pion-nucleon pair production  correspond to the experimental large opening
angle data.
However, the  absolute value of the mechanism contributions to the reactions, conditioned by the isobar
configurations, is by several orders less. Thus, the behavior of the experimental yield of
 the $^{12}$C$( {\gamma ,\pi ^{ -} p})$ reaction for the large opening angles observed  in the
 experiment \cite{10} is impossible to be explained by the effect of isobar configurations in the nucleus ground
 state.

  At present the statistically provided experimental data of the $A( {\gamma ,^{}\pi ^{ +} p})B$
 reaction  in the range of the large momentum transfer to the residual nucleus are absent.
 For  comparison  of  the predictions of the presented photoproduction model with the results of the
 measurements we use the experimental data of the work \cite{30}, in which the cross section
 of the  $^{12}$C$( {\gamma ,^{}\pi ^{ +} p})$ reaction, averaged in  the kinematic area
 with mean value of the residual nucleus momentum equal to $\sim$300 MeV/{\it c}, is measured.
 The average photon energy $\overline{E}_{\gamma}$ was 355\,MeV.

\begin{figure}[t]
\centering
\includegraphics [width = 8cm , keepaspectratio] {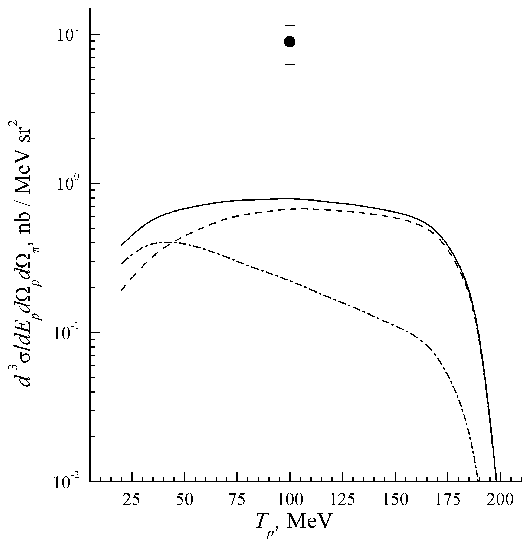}
\vspace*{8pt}
\caption{Differential cross section of the $^{12}$C$( {\gamma ,\pi ^{+} p})$ reaction for
$\overline{E}_{\gamma}= 355$\, MeV versus the kinetic
 energy of the proton $T_p$. Dashed curve: direct reaction mechanism, corresponding to the diagram
 in  Fig. \ref{fig1}\textit{b}, dashed-dotted curve: exchange reaction mechanism, corresponding to
 the diagram in  Fig. \ref{fig2}\textit{e}, solid curve:
the total cross section, data for $\overline{E}_{\gamma}= 355$\, MeV are taken  from Ref. \cite{30}.}
\label{fig5}
\end{figure}

 Fig. \ref{fig5} displays  the results of the calculated cross section  plotted against the kinetic
 energy of the proton  together with the data of the work \cite{30}.
 In Fig. \ref{fig5} the experimental cross section is averaged over the proton energy in the interval
 $80 - 120$ MeV. In addition both experimental and theoretical cross sections  are averaged over the
 pion energy in the interval of
 $71.5 - 106.5$ MeV, and over the proton polar angle in the interval of $56^\circ - 94^\circ$.

  The dashed and dashed-dotted curves
present the cross section contributions of the direct and  exchange reaction mechanisms,
corresponding to the diagram in  Fig. \ref{fig1}\textit{b} and Fig. \ref{fig2}\textit{e}. The contribution
of the exchange mechanism to the reaction, corresponding to the diagram in Fig. \ref{fig2}\textit{b},
 turned out to be less then 10$^{-2}$\,nb/MeV\,{sr}$^{2}$.
According to the used model the probability of the internal excitation of the nucleon
in the nucleus $^{12}$C, as the result of $NN$-interactions, is $\sim$0.01. As it can be seen in Fig. \ref{fig5},
in spite of low probability of the transition $N \to \Delta $, the manifestations of  isobar
configurations in the nucleus ground state  allow to explain some part of the observed
cross section of the reaction $^{12}$C$( {\gamma ,\pi ^{ +} p})$.
One of possible  explanations of the excess of the experimental cross section over the
calculated that is connected with the contribution to the experimental  data of
the $^{12}$C$( {\gamma ,^{}\pi ^{ +} p} )n ^{10}$Be process,
which is realized as a result of the direct  knockout of the correlated
$\Delta ^{ + +} n$ and $\Delta^{+}p$ pairs by photon.

It should be kept in mind that our quantitative conclusion depends on the constants used in model. One
of the uncertainties is connected with the $\Delta^{++}$ magnetic moment. It is necessary to remark that
experimental results and modern theoretical calculations give the $\Delta^{++} $ magnetic moment in
the interval  $(4 - 5)$ of nuclear magnetons. Our estimations show that the cross section uncertainty
introduced by the uncertainty in magnetic moment of the $\Delta^{++}$ isobar is of 30$\%$.

\section{Conclusion}

We considered the production of the pion-nucleon pairs when the  high energy photon
interacted with the nucleus.  We used the model in which the nucleus contains excited states
of the nucleons -- the virtual isobars along with nucleons. The wave function of the  $\Delta $-isobar
configuration in the closed shell nuclei was obtained in the harmonic oscillator model of the nuclei
with the {\it jj}-coupling by means of the solution of the Schrodinger equation. The transition
potential with $\pi$- and $\rho$-exchange was used.

Using the {\it S}-matrix approach, one-particle operator of $\gamma \Delta\to \pi N$ transition was found.
The {\it S}-matrix has been written as the standard expansion in the power of the
interaction Lagrangian neglecting terms above the second order. We have taken into account only the Lagrangian
of the strong interaction of the nucleon, isobar and the pion fields and Lagrangian of the electromagnetic
interaction.
At determination of the transition operator, we took into account  the contributions  from the intermediate
single-particle states  with the smallest mass -- a pion, a nucleon and a $\Delta$(1232)-isobar in
$\textit{s}$-, $\textit{t}$- and $\textit{u}$-channels  of  $\textit{T}$-product  of the currents.

The analysis of the nucleus density matrixes  of the $A(\gamma,\pi N)B$ process  was made.
Direct  and exchange mechanisms of pion-nucleon pair photoprodution which result from one-particle
and two-particle density matrix were considered. The description of the nucleus  as a system
including  alongside with the nucleons  their excitation states brought about the significant
increase of the possible reaction mechanisms set.

We performed calculations of the differential cross section of the \linebreak
$^{12}$C$( {\gamma ,^{}\pi ^{-} p})^{11}$C and $^{12}$C$({\gamma ,\pi ^{ +} p})^{11}$Be reactions.
The numerical estimations of the cross
section value  are made for mechanisms of the reactions, in which the final free nucleon
can have the sufficiently large momentum,
neglecting the exchange mechanisms, in which the nucleon goes to the free state from the state that is
lower the Fermi level.
In the range of the pion and proton opening angle close to 180$^{\circ}$ the total contribution of the
isobar configuration  cross section of the $^{12}$C$( {\gamma ,^{}\pi ^{ -} p})^{11}$C reaction
is smaller than the cross section observed in experiment \cite{10} of about two order of magnitude.
The calculated  cross section of the $^{12}$C$( {\gamma ,\pi ^{ +} p})^{11}$Be reaction is
several times smaller than the experimental cross section of the work \cite{30}.

\end{document}